\documentclass{jpsj-suppl}
\usepackage{txfonts} 
\usepackage{graphicx}
\usepackage{wrapfig}

\title{Hadron tomography for pion and its gravitational form factors}

\author{ Qin-Tao  \textsc{Song}}

\inst{
Department of Particle and Nuclear Physics, Graduate University for Advanced Studies \\
$\ \ $(SOKENDAI), 1-1, Oho, Tsukuba, Ibaraki, 305-0801, Japan\\

KEK Theory Center, Institute of Particle and Nuclear Studies, 
KEK, \\
$\ \ $1-1, Oho, Tsukuba, Ibaraki, 305-0801, Japan 
 }

\email{qintao@post.kek.jp}

\recdate{February 1, 2019}

\abst{Generalized parton distributions (GPDs) are three-dimensional structure functions of hadrons, and they can reveal the orbital-angular-momentum contributions to the nucleon spin. Therefore, GPDs
are important for solving the proton spin puzzle.  
The generalized distribution amplitudes (GDAs) are the $s$-$t$ crossed quantities of the GPDs, and the GDAs can be investigated in two-photon process ($\gamma^* \gamma \to h\bar h$) which is accessible at KEKB. 
The pion GDAs were obtained  by analyzing the Belle measurements for $\pi^0 \pi^0$ production in the $e^+ e^-$ collision. 
From the obtained GDAs, the form factors of energy-momentum tenor were calculated for pion in the timelike region. In order to study the gravitational radii for the pion, the form factors of energy-momentum tenor were obtained in the spacelike region by using the dispersion relation. Then, the mass radius was calculated as  0.32 $\sim$ 0.39 fm and the mechanical radius was also estimated for the pion as 0.82 $\sim$ 0.88 fm  by using the spacelike form factors. This is the first finding on gravitational form factors and radii of hadrons from actual experimental measurements. 
In the near future we can expect more precise measurements of  $\gamma^* \gamma \to h\bar h$ as the Belle II started data taking by the higher luminosity Super KEKB,  so that the GDAs of other hadrons could  be studied as well.  }

\kword{GPDs, GDAs, gravitational form factors,  gravitational radii, QCD}

\begin{document}
\maketitle

\section{Introduction}
\vspace{-0.05cm}

The proton spin puzzle was suggested by the European Muon Collaboration in 1988, and subsequent experiments indicate that only 20\% $\sim $ 30\% of the proton spin comes from quark-spin contribution. In order to solve the proton spin puzzle, one needs to know the orbital-angular-momentum and gluon-spin contributions. The Generalized parton distributions  (GPDs) can be studied in the deeply virtual Compton scattering (DVCS), and the orbital-angular-momentum contributions can be disclosed by the GPDs.  The GPDs are important physical quantities of  hadrons, and one can obtain the parton distribution functions by taking the forward limt of  GPDs. 
Moreover, the GPDs also provide a way to investigate the energy-momentum tensor and gravitational form factors of the hadrons, and those topics become popular recently.

The $s$-$t$  crossed channel of the DVCS is the two-photon process $\gamma^* \gamma \to h\bar h$ \cite{Diehl:1998dk, Diehl:2000uv, Polyakov:1998ze,Kawamura:2013wfa,Kumano:2017lhr}. In this process,  the amplitude can be factorized into two parts if $Q^2$ of the virtual photon is large enough.
The hard part is amplitude of $\gamma^* \gamma \to q\bar q$, and the soft part is the generalized distribution amplitude (GDA) of the hadron which is just the amplitude of  $q\bar q \to h\bar h$
The GDAs are the $s$-$t$  crossed quantities of GPDs, therefore,  the GDA studies can be complementary to the GPD studies. The GPDs and GDAs have an advantage to probe the energy-momentum tensor of the hadron, although the gravitational form factors can not be  measured directly by experiment. As for current status of GDA experiments,  the Belle collaboration reported the cross-section measurements of
the two-photon process $\gamma^* \gamma \to \pi^0 \pi^0$ in 2016 \cite{Masuda:2015yoh}, and the pion GDAs were obtained by analyzing the Belle data \cite{Kumano:2017lhr}.  Furthermore, gravitational form factors and gravitational radii were also estimated. However, the errors of Belle measurements are still large at current stage, and the obtained pion GDA can be affected by the experimental errors.   Belle II began data taking with the much higher luminosity SuperKEKB in 2018,  and the precise measurements of  $\gamma^* \gamma \to h\bar h$ can be expected since the statistic errors are much larger than the  systematic errors in the previous Belle data.

\section{Theoretical formalism for the pion GDA \label{two}}
\vspace{-0.05cm}

\begin{wrapfigure}{r}{0.50\textwidth}
   \vspace{-0.1cm}
   \hspace{0.3cm}
     \includegraphics[width=0.48\textwidth]{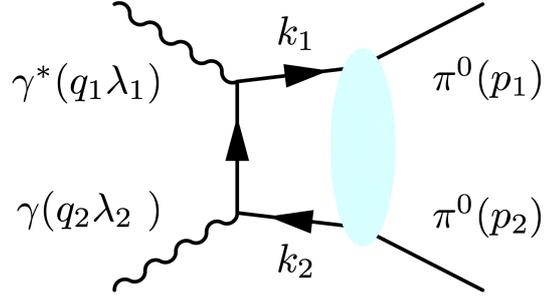}
\vspace{-0.8cm}
    \caption{\label{fig:pion-gda} The two photon process,  and the GDA is just the amplitude of quark-antiquark pair to the hadron-antihadron pair.}
\label{fig:gpd-1}
\vspace{-0.5cm}
\end{wrapfigure}

There are two possible ways to investigate the GDAs. The first one is the two photon process $\gamma^* \gamma \rightarrow h \bar{h}$ 
\cite{Diehl:1998dk, Diehl:2000uv, Polyakov:1998ze}, and this process can be measured in the $e^+ e^-$ collision at KEK B factory. The other possibility is the process  $\gamma^* N  \rightarrow  N h \bar{h}$ \cite{Anikin:2004ja} which can be conducted at JLab and COMPASS.  In the two-photon process, the soft part is the GDA and the hard part can be calculated by perturbative QCD. However,  the amplitude of  $\gamma^* N  \rightarrow  N h \bar{h}$ is determined by both GDAs and GPDs. Therefore, the two photon process is a better choice than $\gamma^* N  \rightarrow  N h \bar{h}$ if one wants to extract the GDAs of hadron.
Here, we take  $\gamma^* \gamma \rightarrow \pi^0 \pi^0$ shown in Fig.\,\ref{fig:pion-gda} as an example to discuss the general propeties of GDAs. In order to satisfy the factorization condition, $Q^2=-q_1^2$ 
should be large enough, namely $Q^2=-q_1^2 \gg W^2=(p_1+p_2)^2$. The pion GDAs describe the amplitude of quark-antiquark pair to the $\pi^0 \pi^0$ pair, which can be expressed as 
\begin{align} 
& \Phi_q^{\pi^0 \pi^0} (z,\zeta,W^2) 
= \int \frac{d y^-}{2\pi}\, e^{i (2z-1)\, P^+ y^- /2}
  \langle \, \pi^0 (p_1) \, \pi^0 (p_2) \, | \, 
 \overline{q}(-y/2) \gamma^+ q(y/2) 
  \, | \, 0 \, \rangle \Big |_{y^+=\vec y_\perp =0} \, ,
\label{eqn:gda-def}
\end{align}
where $P=p_1+p_2$,  the momentum fraction of 
a quark is defined as $z=k_1^+/P^+$,  and $ \xi=p_1^+/P^+$ is the momentum fraction of the pion. The $Q^2$ dependence of $\Phi_q^{\pi^0 \pi^0} (z,\zeta,W^2)$ is abbreviated in Eq. (\ref{eqn:gda-def}).
We denote $A_{\lambda_1 \lambda_2}$ as the helicity amplitude of $\gamma^* \gamma \rightarrow \pi^0 \pi^0$ in Fig.\,\ref{fig:pion-gda},
where $\lambda_1 $ and $\lambda_2$ are the helicities of the virtual photon and the real photon, respectively.
By considering the parity invariance, there are three independent helicity amplitudes $A_{++}$, $A_{0+}$ and $A_{+-}$ \cite{Diehl:1998dk,Diehl:2000uv}.
The amplitude $A_{++}$ is the lead-twist and leading-order contribution with the same helicity $\lambda_1 = \lambda_2$.  The amplitude $A_{0+}$ is the higher-twist effect, and the virtual photon is longitudinal polarized, so it decreases as  $1/Q$.
As for the amplitude $A_{+-}$, it contains the gluon GDA, so it is suppressed by $\alpha_s(Q^2)$ at the high energy \cite{Diehl:2000uv}.

At large $Q^2$, the amplitudes $A_{0+}$ and $A_{+-}$  can be neglected, and the  differential cross section of $\gamma^* \gamma \rightarrow \pi^0 \pi^0$  \cite{Diehl:1998dk,Diehl:2000uv} is expressed as
\begin{align}
d\sigma=\frac{1}{4}  \alpha^2  \pi \frac{\sqrt{1-\frac{4m_\pi^2}{s}} }{ Q^2+s} 
|A_{++}|^2  \sin\theta  d\theta, \quad  A_{++}=\sum_q \frac{e_q^2}{2} \int^1_0 dz \frac{2z-1}{z(1-z)} 
       \Phi^{\pi^0 \pi^0}_q(z, \xi, W^2),
       \label{eqn:amp2}
\end{align}  
where $\alpha$ is the fine structure constant and $\theta$ is the polar angle for $\pi^0$. The amplitude $A_{++}$ is determined by the pion GDA  $\Phi^{\pi^0 \pi^0}_q(z, \xi, W^2)$. According to Eq. (\ref{eqn:amp2}),  one can obtain the pion GDAs by analyzing the cross section measurements of $\gamma^* \gamma \rightarrow \pi^0 \pi^0$.

In the large $Q^2$ limit, the asymptotic form of GDAs can be acquired by solving the ERBL evolution equation \cite{Diehl:2000uv}, and they are independent of $Q^2$.
\begin{align}
&\sum_q \Phi^{\pi^0\pi^0}_q(z, \xi, W^2)=18n_fz(1-z)(2z-1)
[\tilde{B}_{10}(W)+\tilde{B}_{12}(W)P_2(cos\theta)],  
\nonumber \\
&\tilde{B}_{nl}(W)=\bar{B}_{nl}(W) \exp(i\delta_l), \ \ 
\zeta  = \frac{1+\beta \cos\theta}{2}, \ \ 
\beta=\sqrt{1-\frac{4m_\pi^2}{s}}.
\label{eqn:gda}
\end{align}
There are only two terms in Eq. (\ref{eqn:gda}). The term $\tilde{B}_{10}(W)$ 
indicates the   production of S-wave  $\pi^0 \pi^0$, 
and $\tilde{B}_{12}(W)$ is for D-wave $\pi^0 \pi^0$. 
Below the $KK$ threshold, $\delta_0$  is the S-wave phase shift and $\delta_2$ is the D-wave phase shift in the $\pi\pi$ elastic scattering \cite{Bydzovsky:2014cda, Nazari:2016wio}.
Above the  $KK$ threshold, the additional phase is introduced in our analysis.

\begin{wrapfigure}{r}{0.5\textwidth}
   \vspace{-0.1cm}
   \hspace{0.3cm}
     \includegraphics[width=0.48\textwidth]{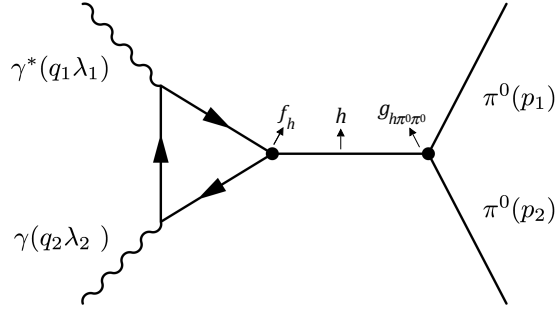}
\vspace{-0.6cm}
    \caption{\label{fig:re}Resonance effect in the two-photon process $\gamma^* \gamma \rightarrow \pi^0 \pi^0$.  }
\label{fig:gpd-1}
\vspace{-0.5cm}
\end{wrapfigure}

In the two-photon process $\gamma^* \gamma \rightarrow \pi^0 \pi^0$, $\pi^0 \pi^0$ can be produced through a intermediate hadron state $h$ in Fig.\,\ref{fig:re}, and this is the so called resonance effect \cite{Anikin:2004ja}.  In Fig.\,\ref{fig:re} the left part is the process of $q \bar{q} \rightarrow h$, and this part can be determined by the distribution amplitude or the decay constant $f_h$ of the hadron $h$. The right part of Fig.\,\ref{fig:re} describes the amplitude $h \rightarrow \pi^0 \pi^0$, and it is related to the coupling constant $g_{h\pi^0 \pi^0}$ which can be obtained with the decay width of  $h \rightarrow \pi^0 \pi^0$. Therefore, resonance contributions for the pion GDAs can be  expressed with the decay constant $f_h$ and the coupling constant $g_{h\pi^0 \pi^0}$. In our analysis, we included the S-wave resonance $f_0(500)$ and the D-wave resonance $f_2(1270)$.

\section{Pion GDA analysis of Belle measurements }
\vspace{-0.05cm}

In 2016, the Belle Collaboration released the measurements of differential cross
section for $\gamma^* \gamma \rightarrow \pi^0 \pi^0$ in the $e^+ e^-$ collision at KEK B factory \cite{Masuda:2015yoh}. Here, we employed the asymptotic form of the GDAs to analyze the Belle data, and resonance effects of $f_0(500)$ and $f_2(1270)$ are introduced in the analysis \cite{Kumano:2017lhr}.
A few parameters of the GDA are determined by fitting the experimental measurements.
\begin{figure}[htpb]
\centering 
\includegraphics[width=0.5\textwidth]{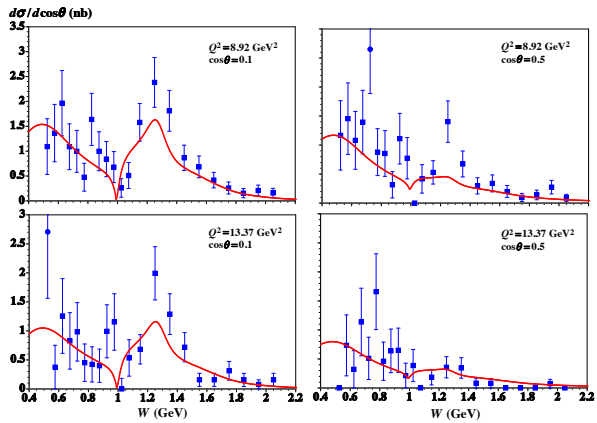}
\caption{\label{fig:aq} The $W$ dependence of the differential cross section (in units of nb),
 in comparison with Belle measurements \cite{Kumano:2017lhr}. } 
\end{figure}
With the obtained GDA, we show the differential cross section of  $\gamma^* \gamma \rightarrow \pi^0 \pi^0$ in comparison with the Belle data in Fig.\,\ref{fig:aq}. The GDA analysis gives a reasonable description of the experimental cross section, and the resonance peak of $f_2(1270)$ is clearly seen around $W=1.27$ GeV. Moreover, one can study the energy-momentum tenor of pion in the timelike region with the obtained GDA, and the matrix element of  energy-momentum tenor is defined as follows
\begin{align}
& \langle \, \pi^0 (p_1) \, \pi^0 (p_2) \, | \sum_q  \, T_q^{\mu\nu} (0) \, 
| \, 0 \, \rangle 
= \frac{1}{2} 
  \left [ \, \left ( s \, g^{\mu\nu} -P^\mu P^\nu \right ) \, \Theta_{1} (s)
                + \Delta^\mu \Delta^\nu \,  \Theta_{2} (s) \,
  \right ],
\label{eqn:emt-ffs-timelike-0}
\end{align}
\begin{wrapfigure}{r}{0.45\textwidth}
   \vspace{-0.1cm}
   \hspace{0.3cm}
     \includegraphics[width=0.44\textwidth]{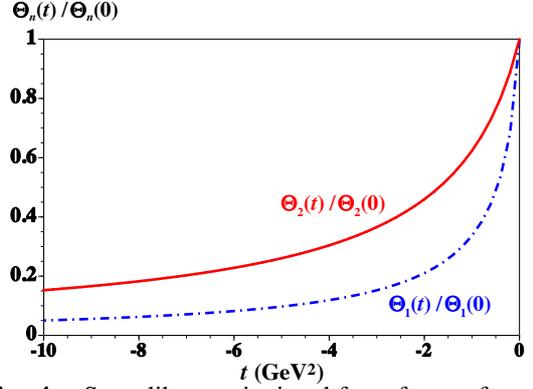}
\vspace{-0.8cm}
    \caption{\label{fig:form}Spacelike gravitational form factors for pion \cite{Kumano:2017lhr}.}
\label{fig:gpd-1}
\vspace{-0.5cm}
\end{wrapfigure}
\noindent where the form factor $\Theta_{1}$ reflects the mechanic properties (pressure and shear force) of hadron , and the form factor $\Theta_{2}$ is related to the mass and energy of hadron.
The gravitational form factors $\Theta_{1}$ and $\Theta_{2}$ are calculated from the pion GDA in the timelike region.
Furthermore, the spacelike gravitational form factors can also be obtained from the timelike ones by using the dispersion relation \cite{form-factor-dispersion}. In Fig.\,\ref{fig:gpd-1}, we show the spacelike gravitational form factors which are normalized as one at $t=0$, and we find $\Theta_{2}$ decreases slower than $\Theta_{1}$ as $|t|$ in creases. 
The mass radius is calculated as $\sqrt {\langle r^2 \rangle _{\text{mass}}} =  0.32 \sim 0.39 \, \text{fm}$ \cite{Kumano:2017lhr} which is just the slope of $\Theta_{2}$ at at $t=0$, and the mass radius is smaller than the charge radius $\sqrt {\langle r^2 \rangle _{\text{charge}}}=0.672 \pm 0.008$ fm 
\cite{Patrignani:2016xqp} of pion.  The range of the mass radius appears since we can introduce the additional phase shift either to the S-wave phase shift $\delta_0$ or to the D-wave phase shift $\delta_2$ above the $KK$ threshold as discussed in  Sec.\, \ref{two}. 
Similarly,  the mechanical radius is obtained as $\sqrt {\langle r^2 \rangle _{\text{mech}}} = 0.82 \sim 0.88 \, \text{fm} $ \cite{Kumano:2017lhr} for $\Theta_{1}$ which is slightly larger than the charge radius.


\section{Summary}
\vspace{-0.05cm}

We  discussed  the basic properties of the GDAs, and the pion GDA was obtained by analyzing the cross section of  $\gamma^* \gamma \rightarrow \pi^0 \pi^0$. 
In the analysis, the resonance effects of $f_0(500)$ and  $f_2(1270)$ were included, and our analysis gave a good description of experimental data. 
The form factors  of energy-momentum tensor were investigated with  the pion GDA  , and gravitational radii were estimated as $\sqrt {\langle r^2 \rangle _{\text{mass}}} =  0.32 \sim 0.39 \, \text{fm}$  for the mass radius and $\sqrt {\langle r^2 \rangle _{\text{mech}}} = 0.82 \sim 0.88 \, \text{fm} $ for the mechanical radius \cite{Kumano:2017lhr}.

\section*{Acknowledgements}
\vspace{-0.4cm}
Q.-T. S is supported by the MEXT Scholarship for foreign students 
through the Graduate University for Advanced Studies.

\end{document}